\documentclass{article}
\usepackage{graphicx} 
\usepackage{titlesec}
\usepackage{float}
\usepackage{url}
\usepackage{hyperref}
\usepackage{authblk}  
\makeatletter

\usepackage{nicefrac}       
\usepackage{lmodern}
\usepackage{microtype}      
\usepackage{booktabs}       
\usepackage{doi}
\usepackage{adjustbox}
\usepackage{float}
\usepackage[T1]{fontenc}
\usepackage{graphicx}
\usepackage{hhline}
\usepackage[utf8]{inputenc}
\usepackage{multicol}
\usepackage{multirow}
\usepackage[normalem]{ulem}

\renewenvironment{abstract}{%
  \if@twocolumn
    \section*{\abstractname}%
  \else
    \small
    \begin{center}%
      {\bfseries \vspace{-2em}}
    \end{center}%
    \quotation
  \fi}
  {\if@twocolumn\else\endquotation\fi}
\makeatother

\title{Neuropsychology of AI:
Relationship Between Activation Proximity and Categorical Proximity Within Neural Categories of Synthetic Cognition
}
\author[1,2]{\textbf{Michael Pichat}}
\author[1,3]{\textbf{Enola Campoli}}
\author[1,4]{\textbf{William Pogrund}}
\author[1,5]{\textbf{Jourdan Wilson}}
\author[1,6]{\textbf{Michael Veillet-Guillem}}
\author[1,7]{\textbf{Anton Melkozerov}}
\author[1,8]{\textbf{Paloma Pichat}}
\author[1]{\textbf{Armanouche Gasparian}}
\author[1,9]{\textbf{Samuel Demarchi}}
\author[1]{\textbf{Judicael Poumay}}

\affil[1]{Neocognition (Chrysippe R\&D) {\href{mailto:contact@neocognition.ai}{contact@neocognition.ai}}}
\affil[2]{University of Paris and Free Faculties of Philosophy and Psychology of Paris (IPC)}
\affil[3]{Department of Cognitive Sciences and Department of Neuropsychology, University of Côte d’Azur}
\affil[4]{Department of Cognitive Sciences, University of Grenoble}
\affil[5]{Department of Linguistics, University of Paris Cité and University of California Los Angeles}
\affil[6]{Epitech Paris}
\affil[7]{Russian Academy of Sciences (FRC CSC RAS)}
\affil[8]{Faculty of Medicine of Lyon East, University of Lyon 1}
\affil[9]{Department of Psychology, University of Paris 8}

\begin{document}

\maketitle
\begin{abstract}
\noindent
Neuropsychology of artificial intelligence focuses on synthetic neural cognition as a new type of study object within cognitive psychology. With the goal of making artificial neural networks of language models more explainable, this approach involves transposing concepts from cognitive psychology to the interpretive construction of artificial neural cognition. The human cognitive concept involved here is categorization, serving as a heuristic for thinking about the process of segmentation and construction of reality carried out by the neural vectors of synthetic cognition.
\end{abstract}

\section{Introduction}

Explainability aims to make the activity of an artificial neural network understandable to humans (Du et al., 2019 ; Pichat, 2023, 2024a). This involves translating the observable behavior of a neural network into an interpretative framework, allowing for the assignment of relevant meaning to this behavior according to the observer's goals. In our context, this framework is that of cognitive psychology. Therefore, it involves using the categories of human cognitive thought as conceptual referents to establish analogies between human and artificial cognitive behaviors. More specifically, we will focus on the notion of categorization, as this concept from cognitive psychology appears particularly relevant for analyzing the synthetic cognition of language models, which largely involves a dynamic extraction of linguistic categorical invariants (Jawahar et al., 2019; Clark et al., 2019; Bills et al., 2023; Clark et al., 2023).

In this work, we focus on an epistemological explainability with fine cognitive granularity (Pichat, 2024b). In other words, we examine a microscopic explainability where the unit of observation is the formal neuron. This low-granularity explanatory approach aims to directly penetrate the "black box" system that an artificial neural network represents by creating elements of understanding about how thought categories and concepts are encoded and structured locally within a language model (Dalvi et al., 2019, 2022). The objective is, therefore, to interpret how categorical knowledge is constructed and utilized by the fundamental elements of the networks, namely the formal neurons themselves (Fan et al., 2023).

\section{Human Categorization}

\subsection{The Cognitive Process of Human Categorization}

Categorization plays a central role in a variety of human cognitive activities of different scales (Sternberg, 2007; Roads et al., 2024): classification and sequencing, identification and denotation of objects, comprehension, reasoning, problem-solving, memorization, inference and prediction, property transfer, conceptualization, etc.

From a formal perspective, a category is defined by two types of elements: its comprehension and its extension (Nadeau, 1999). The comprehension of a category, also called intension, is the set of properties that necessarily and sufficiently define this category, whether these properties are physical, structural, functional, procedural, or goal-oriented (i.e., related to the purpose of the involved task) (Tijus, 2004). Its extension is the set of members belonging to this category.

Historically, the classical notion of a category was shaped by Plato and then Aristotle, who positioned a category as being defined by a series of necessary and sufficient properties. This perspective gave rise, for example, to trait-based categories, conceived as the result of breaking down a category into a series of characteristics, all of which are necessary and collectively sufficient to define that category (Katz, 1972). However, cognitive studies of the actual human processes of categorization quickly revealed the rigidity of this foundational conception of the rules or theoretical defining elements of categorization.

Rosch (1975) developed an approach to categorization by assigning a level of similarity between a candidate object and the prototype of the involved class. This approach is based on the empirical observation that individuals, when asked what defines a category, tend to mention characteristic traits rather than determinative properties (Rosch and Mervis, 1975). The prototype is then defined as the most representative, most typical example of the category, whether it is real or extrapolated through mental construction (such as averaging), in relation to its characteristic traits (Singh et al., 2020; Vogel et al., 2021). This prototype can vary greatly for different subcategories within a given category (Malt and Smith, 1984). In the prototype approach, characteristic traits are frequently present in the items that make up the class, but this is not always the case. Indeed, this approach to categorization, which is more flexible than the previous definition-based conception, addresses Wittgenstein's (1953) "family resemblance" objection: if the prototype is characterized by properties (1, 2, 3, 4), an element A possessing attributes (1, 2, 5, 6) and an element B possessing attributes (7, 8, 3, 4) can both be assigned to the same category even if they do not share any common attributes. This prototype approach is also found in the field of \textit{princeps} perception theories (Posner et al., 1967; Bransford, 1971; Reed, 1972).

Related but different from the prototype approach is the exemplar theory of categorization (Medin and Schaffer, 1978; Nosofsky, 1992; Nosofsky et al., 2022), which suggests that objects are categorized by comparison to typical examples of the category, examples stored in memory. The most typical exemplar is the one that most closely resembles all the known exemplars by the individual. This exemplar exerts the strongest attraction power due to its frequency of occurrence among the exemplars retained at the memory level.

Let us briefly note, even though we will not explore it further, the definition of categorization by semantic networks (Collins and Quillian, 1969; Hornsby et al., 2020). In this conception, categories are structured within a network of nodes (concepts) and links between these nodes (relationships between concepts), ranging from the most specific categories to those with the highest level of generality.

Finally, let us mention the contextual, circumstantial, or goal-oriented approach to categorization, similar to \textit{ad hoc} categories (Barsalou, 1983; Glaser et al., 2020). In the context of these functional approaches (Rips, 1989; Keil, 1989; Wisniewski and Medin, 1994; Barsalou, 1995; Bove et al., 2022), the goal becomes the central element in defining a category, rather than general logic, semantics, or appearance. Here, it is the situation, through its end goal and specific context elements, that guides the categorization, and the terms of this categorization have no existence outside of this situation, \textit{hic et nunc}.

\subsection{Human Categorization by Similarity Judgment}

Similarity-based categorization approaches assume that an object is assimilated to a class based on its estimated proximity to what represents the class (Thibault, 1997; Jacob et al., 2021; Kaniuth et al., 2022; Roads et al., 2021, 2024). This is based on (i) a space of traits or dimensions deemed relevant for comparison and (ii) a mode of calculating the distance between compared instances.

The involvement of similarity judgment as a basis for categorization seems broad-spectrum (Thibault, 1997), especially for classes without explicit definitions and with hierarchical organization, making it possible for some items to clearly belong to a category (Hampton, 1997).

The theories of categorization by prototype (Posner and Keele, 1968; Reed, 1972; Rosch \& Mervis, 1975; Medin and Schaffer, 1978) mentioned previously, in fact, highlight the role of similarity in the categorization process (Sanborn et al., 2021): an item is assigned to a category if it is judged to be close to the central representation, which is the prototype. The same applies to the exemplar-based approach to categorization (Medin and Schaffer, 1978; Brooks, 1987; Nosofsky, 1992): an item is assigned to a categorical class if it is estimated to be closest to the significant elements that make up that class. In both cases, categorization results from the estimated distance between the item in question and what represents the category (Ayeldeen et al., 2015; Roads et al., 2024).

\subsection[Arguments Against Human Categorization by Similarity]{Arguments Against Human Categorization by \\ Similarity}

Arguments against the effective or possible foundation of categorization on similarity reasoning are diverse (Love, 2002):
\begin{itemize}
    \item An element is assigned to the category that best explains it (Murphy and Medin, 1985), beyond possible initial classification by similarity (Keil, 1989).
    \item Categories with explicit definitions cannot be directly based on similarity judgment (Kalyan et al., 2012).
\end{itemize}

However, the main arguments are based on the idea that the singular choice of similarity judgment criteria, which are just one possibility among others in the space of traits or dimensions, does not necessarily align with what constitutes or should constitute categorical assignment (Reppa et al., 2013; Poth, 2023):
\begin{itemize}
    \item Categorization is influenced by information external to the objects being classified: general theories about the world or elements specifically related to the category involved (Rips, 1989).
    \item The assignment of an object to a category is also a function of the relationships between the other objects that make up that category (Medin et al., 1993).
    \item In cases where categorization is determined by the purpose of tasks, particularly in the case of \textit{ad hoc} categories (Barsalou, 1991), similarity will invoke comparison criteria that are not suited to this finalized activity.
\end{itemize}
These arguments converge on the idea that reasoning based on similarity is too ambiguous to functionally underpin categorical assignment (Wixted, 2018). Rips (1989) mentions in this regard a non-monotonic relationship between similarity and category membership. The criterion used for a similarity judgment indeed varies depending on the context (Murphy and Medin, 1985), for example, cultural contexts (Whorf, 1941). In other words, the criteria used for similarity judgment are not sufficiently constrained and are therefore too dependent on the singular choice of segmentation made \textit{hic et nunc} (Goodman, 1972).

Thus, it follows that similarity judgments and category membership are not in alignment (Rips, 1989; Medin, 1993) and that similarity cannot, or should not, imply categorization.

\subsection{Counterarguments in Favor of Human Categorization Based on Similarity}

In response to arguments against the involvement or relevance of similarity in the categorization process, several responses have been provided (Bobadilla et al., 2020; Hebart et al., 2020).

Goldstone (1994) offers the following fourfold counter-argument: (i) The argument that similarity is too unstable to serve as a basis for categorization does not hold because it presupposes that categorization itself would not also be flexible; (ii) Even if superficial in some cases, similarity is functional in that it can genetically facilitate the discovery of "deeper" indicators of categorization and thereby the creation of new, more "fundamental" categories; (iii) Experiments show that similarity is not as unstable as is often argued; (iv) Categories that are not based on similarity are resistant to generalization.

Thibault (1997), in response to criticisms about the subjective relativity of similarity, posits that categorization is actually a sub-type of similarity. The author acknowledges that while similarity is indeed contingent upon the choice of comparison criteria, categorization operates similarly, except that it selects its own criteria from a set of traits defining the relevant category. Criticizing psychological essentialism, Thibault (idem) also asserts that the argument regarding the weakness of similarity (in the face of contextual elements, for example) does not hold, as this position assumes that the criteria for categorical segmentation have, could have, or should have an intrinsic, ontological \textit{per se} value, independent of the individual.

Finally, Hampton (1997) argues that categorization itself can be affected by irrelevant similarity elements (often perceptual and notably visual), or at least those deemed irrelevant by \textit{a priori} logical analysis that dogmatically asserts that thought should operate in accordance with the standards of what is instantiated as logic. Moreover, the author shows, based on a fuzzy logic approach, that while subjects sometimes find it difficult to define categories, they nevertheless have no difficulty indicating the extent to which two items from these categories differ or identifying typical members, i.e., engaging in cognitive activities of similarity regarding them. In both cases, once again, Hampton emphasizes that the cognitive weaknesses attributed to similarity are based on an erroneous assumption of the subordination of categorization to what is normatively and \textit{a priori} instantiated as \textit{the} (classical) Logic.

An invariant emerges from these counter-arguments: asserting the limits of similarity as a basis for categorization involves a dual epistemological flaw: (i) A realist flaw, which involves artificially decreeing categorization as a process that engages with or must engage with a reality that is ontologically predefined; (ii) A rationalist flaw, which assigns to categorization a (duty of) subordination to a logic invoked as being self-evident, a self-evidence that every individual should also be able to grasp.

\section{Problem Statement}
\subsection{Context}

Numerous studies reveal or infer a diversity of categories (linguistic, logical, positional, etc.) encoded in neurons and attention heads. In the classic experiment by Clark et al. (2019) on BERT, the authors highlight the converging linguistic functions of attention heads from the same layers. In their fascinating study on GPT-2XL, Bills et al. (2023) identify a series of specific neurons, noting that some are highly sensitive to context. Research also shows a geographic distribution of the type of categorial neural activity according to layer depth. Thus, the early layers respond more to morphological categories at the word level, while the deeper layers are more sensitive to the syntactic categorical features of sentences (such as passive/active voice, tense) and semantic categories (Jawahar et al., 2019).

In the context of our present work, we build on certain aspects of the study conducted by Bills et al. (2023), redirecting them toward other, more cognitive and epistemological issues (Pichat, 2024), as we will specify in the following section. Before that, let us briefly explain the approach of Bills et al. (idem). Based on the hypothesis that a neuron activates specifically for a given property, the researchers undertook a comprehensive analysis of the categorical semantics of all the neurons in GPT-2XL. Methodologically, they subjected GPT-2XL to an extensive series of token sequences, randomly selected from the internet data used to train the model. For each token, the activation values of all neurons across all layers were recorded. GPT-4 was then used to automatically identify the elements to which each neuron reacts (i.e., to generate the "categorical explanation") based on an instructional and example prompt applied only to the five sequences with the highest activations.

\subsection{From Human to Synthetic Categorization}

In line with the question we have presented in the field of human cognition regarding the relationship between categorization and similarity, our transposition of this heuristic question into the field of synthetic cognition is as follows: Is the level of categorical membership of tokens (arriving at a given neuron in the form of embeddings) to the category associated with that neuron related to their level of similarity? In other words, are the intensity of categorical membership and the intensity of similarity of tokens, as analyzed by a given neuron, two related aspects of the same phenomenon? Put differently, is the neural space of categorical membership segmented according to the segmentation of the similarity space? This is a largely unexplored issue in the field of synthetic explainability (Fan et al., 2023; Luo et al., 2024; Zhao et al., 2024), and it seems particularly pertinent to investigate.

From an epistemological perspective, we have transposed the notion of categorical membership (measured on a dichotomous nominal scale, yes/no) in the realm of human cognition to the notion of the level of categorical membership (thus measured on an ordinal scale) in the realm of synthetic cognition; this is because, in the artificial neural field, categorical membership (i.e., activation, as we will elaborate later) is a numerical value rather than a boolean one.

\section{Methodology}

\subsection{Choice of GPT-2XL}

In our study, we focused on OpenAI's GPT models because this suite of models, which inaugurated a significant part of contemporary generative AI, presents the paradox of being both the most popular in terms of media coverage and user numbers, while at the same time being the least studied directly internally, that is, in terms of fine-granularity explainability. The model chosen is GPT-2XL. This model is of particular interest because it is sophisticated enough to study the high-level synthetic cognitive phenomena that interest us without reaching the complexity of GPT-4, and even more so the multimodal GPT-4o, whose complexity does not seem appropriate for the initial cognitive inquiry we are pursuing; in other words, GPT-2XL appears to offer a good level of compromise. Beyond this epistemic reason, a pragmatic reason also guided our choice of GPT-2XL: for the first time, in 2023, OpenAI broke its "black box" tradition (which is certainly logical from a commercial standpoint) regarding its products by providing, in the context of the article by Bills et al. (2023), information on the parameters and activation values of the neurons constituting GPT-2XL. These parameters and activation values will therefore serve as our starting data for this study.

For all practical purposes, let us clarify that GPT-2XL is the broad-spectrum variant of GPT (Generative Pre-trained Transformer) 2, developed by OpenAI and released in 2019. As its name suggests, GPT is a transformer, combining layers of attention heads and feed-forward perceptron-type layers. Its activation function is GeLU. Resulting from unsupervised training (at least directly) on a dataset of 8 million web pages, the model has approximately 1.5 billion parameters distributed across 48 layers. Each of these layers consists of 6.400 neurons and operates on 1.600-dimensional embeddings; each layer (or transformer block) comprises an attention sub-layer with 25 attention heads and two feed-forward network sub-layers. The purpose of the model's training is text completion and generation, making it capable, within its performance range, of a variety of tasks.

\subsection{Our Specific Data Choices}

For the sake of simplification, in this exploratory study, we limited ourselves to the first two layers of GPT-2XL (layer 0 and layer 1) and the 6.400 neurons in each of these layers.

Regarding the tokens and their activation values within these 2 x 6.400 = 12.800 formal neurons, we chose, for each of these neurons, to consider as relevant data its top 100 most activated tokens on average, along with their respective activation values. Indeed, the selection of only hyperactivated tokens, as conducted by Bills et al. (2023), seems too restrictive to us because it is not representative of the variability of tokens for which a neuron activates, potentially giving us a too limited view of the category of tokens to which a given neuron responds. In other words, we believe that Bills et al. (idem) may not interpret neurons extensively but instead identify a very limited subcategory of the category encoded by each of these neurons. Another argument that informed our choice is that the tokens with high average activation values selected are, \textit{de facto}, less sensitive to contextual effects, which, although crucial (and this constitutes, in a sense, a limitation of our approach), can themselves also limit the extension of tokens belonging to a given neural category and thus the categorical semantics of the neurons involved.

\subsection{The Interpretative Construction of Our Observables}

As we just said before, the average activation level of a token within a neuron appears to us as a good operationalization of the equivalent in synthetic cognition of the level of categorical membership in human cognition. Indeed, the average activation of the 100 most activated tokens seems to us to be effectively representative of the extent to which these tokens are part of the extension of a category. This is, in the field of synthetic cognition, in line with the hypothesis of Bills et al. (2023) that a neuron activates specifically for a property, and the foundations of low-granularity explainability studies. Let us illustrate this idea with a case highlighted by Bills et al. (idem): concerning neurons that activate after a repeated occurrence of tokens, the stronger the repetition (i.e., the more the involved token sequence meets the defining condition of this category, namely, token repetition), the stronger the activation. This transposition between the level of categorical membership and the level of activation also seems justified, this time in the neurobiological field of human cognition, by the fact that the activation function is the analog corollary of the transfer function (Savioz et al., 2010), whose purpose is precisely to clarify the inputs belonging to the category to which a biological neuron should react, by increasing the signal-to-noise contrast, figure/ground (Servan-Schreiber, 1990), that is, the contrast between what belongs to the category of elements for which the neuron should activate versus residual elements.

Similarly, the cosine similarity between two tokens appears to us as a good measure of the notion of similarity between two items transposed from the domain of human cognition to that of artificial cognition. Indeed, in the domain of human thought, as we have seen, Thibault (1997) defines similarity based on (i) a space of traits or dimensions considered relevant for making the comparison and (ii) a mode of calculating the distance between the compared instances. In accordance with this definition, cosine similarity is a commonly used measure in NLP to gauge semantic proximity between two elements (Ham, 2023); this is based on the dot product between the two involved multidimensional vectors, which consists of measuring the distance, dimension by dimension in the semantic vector space at play, between the two items to be compared. Cosine similarity can be normalized from -1 to 1, where -1 indicates opposite vectors (opposite similarity), 0 indicates orthogonal vectors (no similarity), and +1 indicates identical vectors (total similarity). It should be noted that we made the central choice to measure cosine similarity within the embedding base of GPT-2XL, and not, for example, in the more performant base of GPT-4, to avoid the methodological limitation mentioned by Bills et al. (2023) and Bricken (2023) of matching synthetic cognitive systems that do not rely on the same embedding system, i.e., not on the same categorical segmentation display. However, for purposes of comparison and plausibility checks of our data, we have also systematically used three other freely available classic embedding bases: Alibaba-NLP/gte-large-en-v1.5, Mixedbread-ai/mxbai-embed-large-v1, and WhereIsAI/UAE-Large-V1.

\subsection{Statistical Details}

Our descriptive and inferential statistical calculations were conducted using Python libraries from the SciPy suite, based on the guidelines of Howell (2008) and Beaufils (1996).

The preliminary study of data normality, conducted for exploratory purposes as well as to verify the conditions for conducting certain parametric tests (ANOVA, regression, Grubbs' test), was twofold. Firstly, it was done using various inferential tests, each with its respective advantages: Shapiro-Wilk test (valid for small samples), Lilliefors test (valid for small samples and cases where the parameters of the normal distribution are unknown and estimated from the data), Kolmogorov-Smirnov test (better suited for large samples), and Jarque-Bera test (focused on skewness and kurtosis, valid for large samples). It should be noted that, concerning the cosine similarity measurements, these tests were systematically performed on all four mentioned embeddings to verify the convergence of the results. Secondly, a numerical descriptive approach was employed (skewness and kurtosis indices, mean-median deviation) as well as a graphical approach (QQ-plot comparing the actual distribution with the theoretical normal distribution). This variety of approaches provides us with a broad-spectrum view.

Two types of statistical units were identified. For our "micro" investigations, neuron by neuron, the instantiated statistical units are tokens, specifically the 100 most activated tokens on average for each neuron, which we call "core tokens." On these first-order statistical units, the following parametric tests were conducted: Fisher's test for linear regression with normally distributed residuals (comparing the variance explained by the regression model with the unexplained variance), Grubbs' test for identifying outliers in a normal distribution, and the univariate Student's t-test for comparing a mean to a standard on a normally distributed variable (with correction for small sample sizes). The following non-parametric tests were also conducted: Spearman's rhô (for correlations on an ordinal scale) and Wilcoxon-Mann-Whitney (for comparing group means on an ordinal scale). Regarding the cosine similarity measures, all these parametric and non-parametric tests were performed on the four indicated embeddings bases, again to verify the convergence of our results.

For the "macro" investigations, which provide a global inference across all neurons of a given layer, the statistical units considered were neurons, specifically the 6.400 neurons of each layer. On these second-order statistical units, only one non-parametric test was performed (primarily on the GPT-2XL embedding): the univariate \(\chi^2\) goodness-of-fit test , allowing us to infer the significance of "micro" phenomena on the entire "macro" scale of neurons in a given layer. It should be noted that if our study had been broader, covering all 48 layers of GPT-2XL, a third relevant type of statistical unit would have emerged, that of the layers, allowing for the generalization of identified phenomena to the entire model.

\subsection{The Question Investigated and Its Operationalization}

By choosing, as previously mentioned, to operationalize categorical membership through activation and similarity through the cosine similarity measure, our operationalized initial question becomes: Is the activation level of tokens related to their cosine similarity level? Expressed in functional terms (in the mathematical sense), this question is formulated as follows: Is there a relationship between activation (in the activation space) and cosine similarity distance (in the cosine similarity space)?

To further operationalize this question, we choose to study it from the perspective of the proximity of activation intensity between tokens. This is based on the inference that if the level of similarity (i.e., categorical proximity) were related to the level of categorical membership, it would likely imply a relationship between the proximity of categorical similarity intensity and the proximity of categorical membership intensity. The question then becomes operationalized as follows: Is there a relationship between activation proximity and cosine similarity (proximity) between tokens?

In terms of statistical units, as already indicated, we choose to focus, for each neuron, on the 100 tokens that are subject to its 100 highest average activations. This choice is also made because it does not seem relevant to us to focus on tokens that do not belong, or belong only weakly, to the category associated with each neuron (i.e., those that are not highly activated); indeed, it seems unlikely, except by statistical chance, to find such tokens that would be systematically linked in terms of similarity. Thus, our question, as instantiated concerning activation proximity, guides us methodologically towards the following final choice of primary statistical units: the series of pairs of successive core tokens concerning their level of activation.

Ultimately, our question becomes: Is there, at the level of successive core-token pairs for each neuron, a relationship between activation proximity and cosine similarity (proximity)?

\section{Results}

\subsection[Preliminary Statistical Explorations for Methodological Purposes]{Preliminary Statistical Explorations for \\ Methodological Purposes}

For both layers (see Tables No. 1 \& 2 and corresponding Graphs No. 1 \& 2), the comparison of minimum, mean, range, \(Q_1\), and CV (coefficient of variation) values for the cosine similarity between successive core-token pairs (as per their activation level) appears to highlight a relative deficiency in the three embedding models—Alibaba, Mixedbred, and WhereIsAI—compared to GPT-2XL: the latter exhibits a greater discriminative power. This phenomenon could potentially be partially explained by the following methodological bias, as intentionally noted earlier in our methodological section: the core tokens involved are \textit{de facto} compliant with the GPT-2XL tokenization system and not necessarily aligned with the segmentation modalities that governed the other three embedding databases. As a result, in all cases, the cosine values from the GPT-2XL embeddings will be considered more reliable in the continuation of this study. However, this does not imply outright rejection of insights from the other embedding models for the purpose of (i) verifying the inter-embedding convergence of our results, and (ii) especially since the current results are highly convergent among these three embedding systems, which would still argue for a certain reliability concerning them.

\begin{table}[!htbp]
\renewcommand{\arraystretch}{1.3}
\begin{adjustbox}{max width=\textwidth}
\begin{tabular}{p{0.48\textwidth}p{0.48\textwidth}} 
\includegraphics[height=3.2cm]{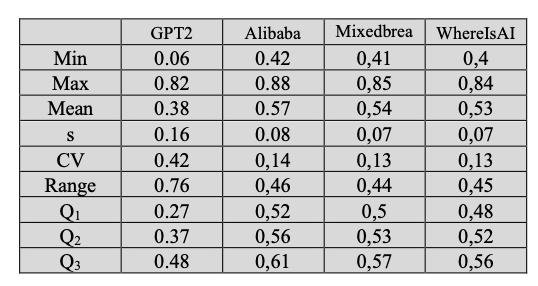}
\newline
\raggedright
\parbox{\linewidth}{\centering
\footnotesize
\textit{Table 1  : Statistical averages of the descriptive indices of position and dispersion for cosine similarities of pairs of successive core-tokens (Layer 0, n = 6400).}} &
\includegraphics[height=3.1cm]{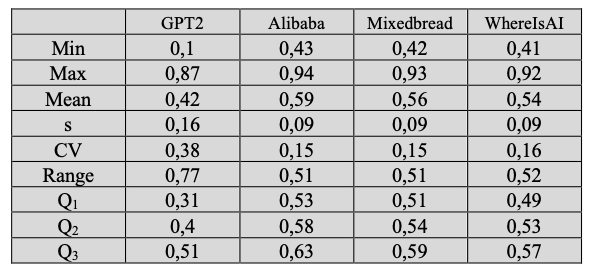}
\newline
\raggedright
\parbox{\linewidth}{\centering
\footnotesize
\textit{Table 2 : Statistical averages of the descriptive indices of position and dispersion for cosine similarities of pairs of successive core-tokens (Layer 1, n = 6400).}} \\
\end{tabular}
\end{adjustbox}
\end{table}

\begin{table}[H]  
\renewcommand{\arraystretch}{1.3}
\begin{adjustbox}{max width=\textwidth}
\begin{tabular}{p{0.48\textwidth}p{0.48\textwidth}} 
\includegraphics[height=2cm]{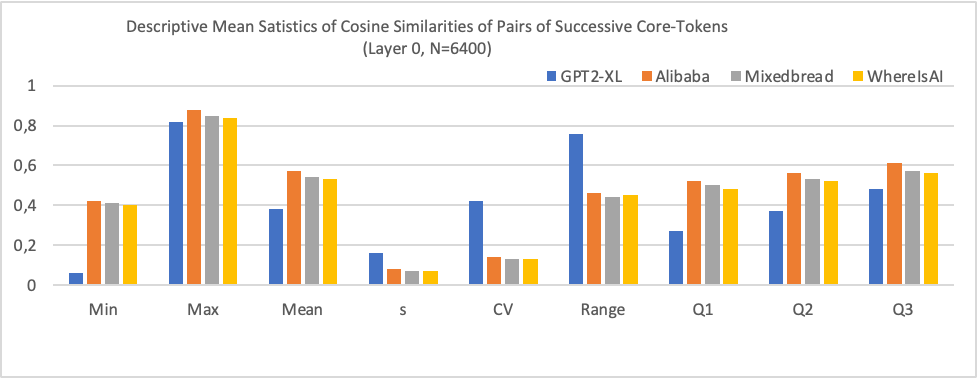}
\newline
\raggedright
\parbox{\linewidth}{\centering
\footnotesize
\textit{Graph 1: Statistical Averages of Descriptive Indices of Position and Dispersion of Cosine Similarity for Pairs of Successive Core-Tokens (Layer 0, n=6400).}} &
\includegraphics[height=2cm]{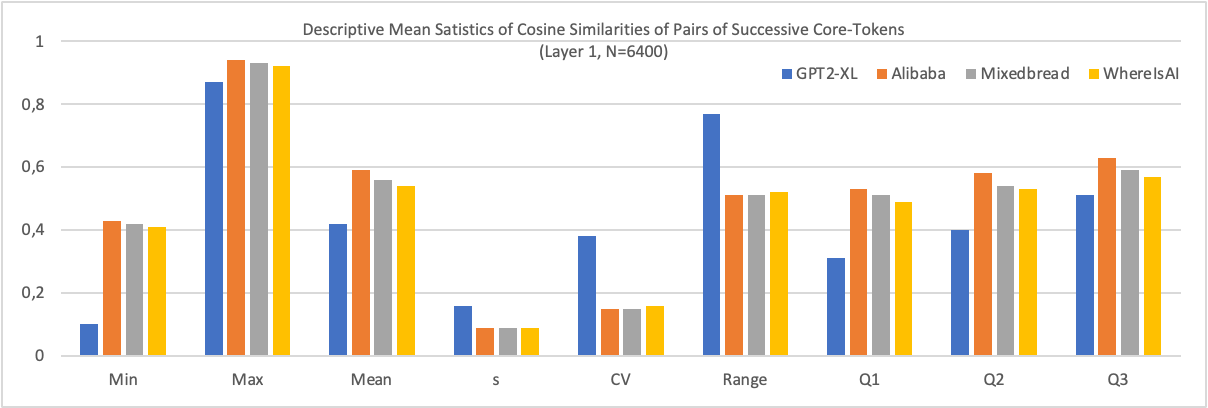}
\newline
\raggedright
\parbox{\linewidth}{\centering
\footnotesize
\textit{Graph 2: Statistical Averages of Descriptive Indices of Position and Dispersion of Cosine Similarity for Pairs of Successive Core-Tokens (Layer 1, n=6400).}} \\
\end{tabular}
\end{adjustbox}
\end{table}


For both layers (refer to Tables No. 3 \& 4), we observe that inferential indicators (Shapiro-Wilk, Lilliefors, Kolmogorov-Smirnov, and Jarque-Bera; for \(\alpha = .05\)) as well as descriptive ones (comparison mean/mode, symmetry, and kurtosis) related to the cosine similarity between successive core-token pairs are compatible with (which does not necessarily prove) a normality hypothesis in 2/3 of cases for measurements made from the GPT-2XL embeddings. However, these indicators consistently drop for measurements based on the other three embedding systems (see appendices for results related to "control neurons", including QQ-plots / Henry's line graphs based solely on GPT-2XL embeddings). This divergence may again be partially explained by the variability of tokenization systems. In both cases, normality appears lesser for layer 1 compared to layer 0. Nonetheless, these results suggest greater relevance, in our upcoming statistical setups, for the use of non-parametric tests (i.e., not assuming normality of the variable involved), or at least particular caution in interpreting some of our partial results that will be based on parametric tests.

\begin{table}[!htbp]
\renewcommand{\arraystretch}{1.3}
\begin{adjustbox}{max width=\textwidth}
\begin{tabular}{p{0.48\textwidth}p{0.48\textwidth}} 
\includegraphics[height=3cm]{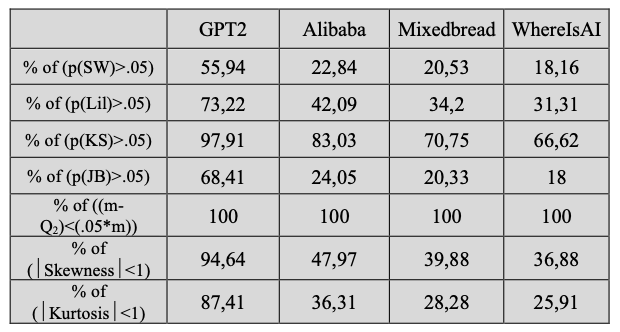}
\newline
\raggedright
\parbox{\linewidth}{\centering
\footnotesize
\textit{Table 3 : Percentages of inferential statistics (\(\alpha = .05\)) and descriptive normality of cosine similarity of pairs of successive core-tokens (Layer 0, n=6400).}} &
\includegraphics[height=3cm]{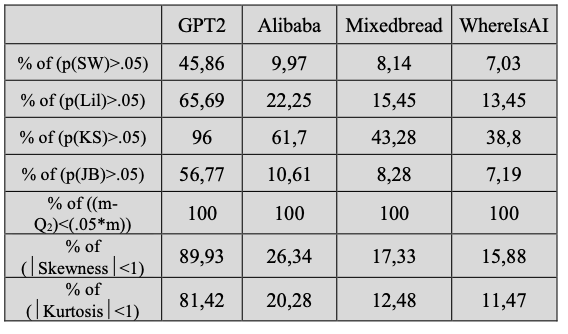}
\newline
\raggedright
\parbox{\linewidth}{\centering
\footnotesize
\textit{Table 4 : Percentages of inferential statistics (\(\alpha = .05\)) and descriptive normality of cosine similarity of pairs of successive core-tokens (Layer 1, n=6400).}} \\
\end{tabular}
\end{adjustbox}
\end{table}

\subsection{Activation Proximity and Cosine Proximity}
 
In our investigation of the relationship between activation proximity and cosine (similarity) proximity among successive core-tokens of each neuron, Tables No. 1 (neurons from layer 0) and No. 2 (neurons from layer 1) seem enlightening. We indeed observe low averages, relative to a theoretical positive span from 0 to 1 of cosine similarity values, with respective values of .38 and .42 for measurements made based on the embeddings from GPT-2XL. An inferential \(\chi^2\) goodness-of-fit measurement conducted on the percentage of neurons having average cosine similarity values below .5 (i.e., relatively low), assuming a theoretical equidistribution hypothesis, is fully compatible with this initial descriptive observation (\(p(\chi^2) < .05\) for both layers, N=6400). This result is also consistent with the similarly low average \(Q_3\) values (respectively .48 and .51), and the very low average minimum cosine values (respectively .06 and .1). This exploratory view supports the notion, at the level of the overall distribution of cosines taken as a whole, that activation proximity (i.e., categorical level closeness between two tokens) does not coincide with cosine proximity (i.e., categorical proximity between these two tokens).

Graphs No. 3 (control neuron 0 from layer 0) and No. 4 (control neuron 0 from layer 1) show representative examples of the distribution of cosine similarity as a function of the activation value of the first token in each pair (for the 100 core-tokens selected) (see appendices for other control neurons). From these specific examples, we can globally observe: (i) again relatively low values of cosine similarity (notably in the case of measurements with GPT-2XL embeddings), (ii) a qualitatively significant variability of cosine similarity; this being consistent across the four embedding models (even though the variability seems more pronounced when the cosine is calculated from GPT-2XL embeddings, which is expected given the more discriminant semantic power of this embedding system for our specific data). This qualitative view, as it only pertains to examples, illustrates the potential fact that, at the level of the overall distribution of cosines, activation proximity does not correlate with cosine proximity; indeed, we do not obtain here stable graphs (i.e., linear of the type y=a) with a relatively high and constant value of cosine similarity for successive core-tokens regarding their activation level. 

\begin{table}[H]  
\renewcommand{\arraystretch}{1.3}
\begin{adjustbox}{max width=\textwidth}
\begin{tabular}{p{0.48\textwidth}p{0.48\textwidth}} 
\includegraphics[width=\linewidth]{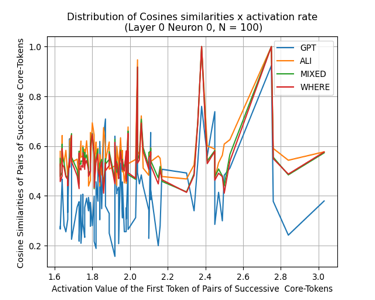}
\newline
\raggedright
\parbox{\linewidth}{\centering
\footnotesize
\textit{Graph 3: Distribution of Cosine Similarity Between Pairs of Successive Core-Tokens as a Function of the Activation Value of the First Token in Each Pair (Layer 0, Neuron 0).}} &
\includegraphics[width=\linewidth]{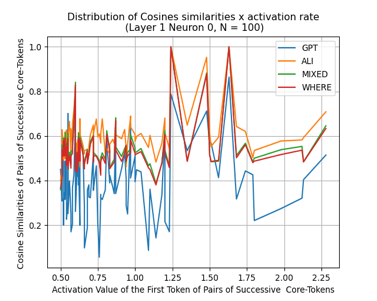}
\newline
\raggedright
\parbox{\linewidth}{\centering
\footnotesize
\textit{Graph 4: Distribution of Cosine Similarity Between Pairs of Successive Core-Tokens as a Function of the Activation Value of the First Token in Each Pair (Layer 1, Neuron 0).}} \\
\end{tabular}
\end{adjustbox}
\end{table}


We will next quantitatively study more specifically this initial global trend of metric non-equivalence between activation proximity and cosine proximity; through two phenomena of synthetic cognition, aimed at exploring an extreme version of this trend, to demonstrate the potential strength it may possess.

\subsection{Categorical Discontinuity of Successive Core-Tokens}

To investigate the previously mentioned trend, namely that, at the level of the overall distribution of cosines, activation proximity (i.e., the level of categorical belongingness between two tokens) does not equate to cosine proximity (i.e., categorical proximity between these two tokens), we test the following initial hypothesis, which represents a first perspective on this trend, extremized as already indicated to demonstrate the potential intensity it might occasionally exhibit: there is a categorical discontinuity in successive core-tokens regarding their level of activation. In other words, there are categorical breakpoints (i.e., semantic breaks) between successive core-tokens. Put another way, there are particularly low cosine similarities between successive core-tokens relative to their level of activation.

To test this, we first operationalize this hypothesis in terms of outliers in the distribution of the cosine similarity variable, more precisely lower outliers, with the notion of a lower outlier perfectly embodying the spirit of our hypothesis. Tables No. 5 (layer 0) and No. 6 (layer 1) show the average numbers of significant (\(p < .05\)) lower outliers per neuron obtained inferentially with the Grubbs test. These average numbers (respectively .007 and .005 with GPT-2XL embeddings) are very low and do not support our hypothesis. However, the reliability of this test is questionable here since the condition for its application, the normality of the distribution of the cosine similarity variable, is not well verified as previously indicated. A non-parametric approach, here the interquartile range, is therefore more secure; and it shows more lower outliers: on average, respectively .151 and .149 lower outliers per neuron (again with GPT-2XL), with extremely low average cosine similarity means (respectively .057 and .082), clearly demonstrating the strong intensity that the categorical discontinuity of successive core-tokens can sometimes take, even if this phenomenon remains here marginal (but not nonexistent) and relatively statistically normal when operationalized with an outliers approach. Table No. 7 qualitatively illustrates this categorical discontinuity by showing core-tokens that are semantically quite distant from each other.

\begin{table}[H]  
\renewcommand{\arraystretch}{1.3}
\begin{adjustbox}{max width=\textwidth}
\begin{tabular}{p{0.48\textwidth}p{0.48\textwidth}} 
\includegraphics[width=\linewidth]{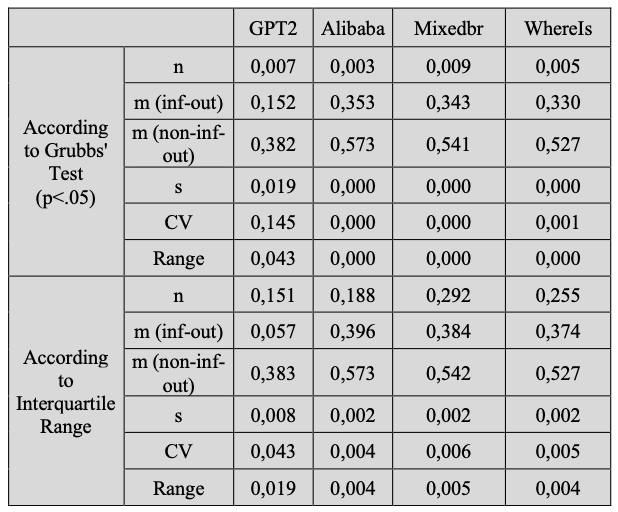}
\newline
\raggedright
\parbox{\linewidth}{\centering
\footnotesize
\textit{Table 5 : Average statistics of lower outliers of cosine similarity for successive core-token pairs (Layer 0, n=6400).}} &
\includegraphics[width=\linewidth]{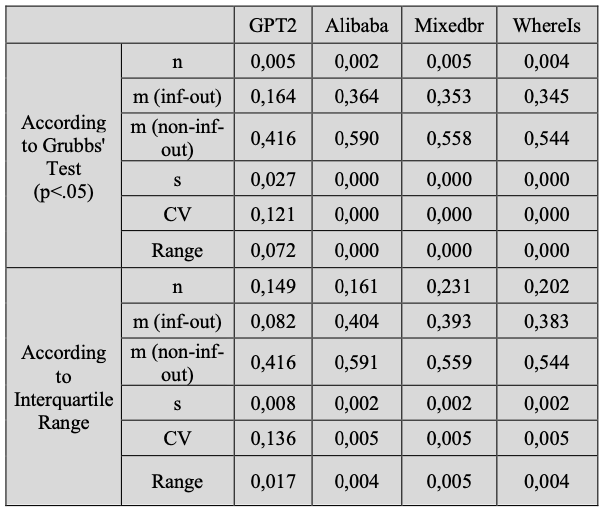}
\newline
\raggedright
\parbox{\linewidth}{\centering
\footnotesize
\textit{Table 6 : Average statistics of lower outliers of cosine similarity for successive core-token pairs (Layer 1, n=6400).}} \\
\end{tabular}
\end{adjustbox}
\end{table}

\begin{table}[H]  
\renewcommand{\arraystretch}{1.3}\centering
\includegraphics[width=\linewidth]{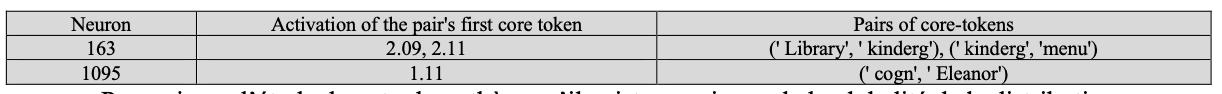}
\newline
\raggedright
\parbox{\linewidth}{\centering
\footnotesize
\textit{Table 7 : Examples of lower outliers cosine similarity for successive core-token pairs (Layer 1, calculated using interquartile range on GPT-2XL, embeddings).}} 
\newline
\end{table}

We continue to study our hypothesis that there exists, at the level of the overall distribution of cosines, particularly low cosine similarities between successive core-tokens relative to their level of activation; but this time, with a less extremized operationalization, intended to make it more manifest. This is done by taking as an indicator the low cosines, which we define, neuron by neuron, as being lower than the threshold of the minimum cosine of the neuron increased by 10\% of its range; this corresponds, in average values, to thresholds of .14 for neuron 0 and .18 for neuron 1 (cf. Tables No. 1 \& 2). We then mechanically observe (cf. Tables No. 8 \& 9) significantly higher average percentages of frequencies of low values per neuron (measured with GPT-2XL embeddings), respectively 5.06\% and 5.17\%; this, with very low cosine averages (.089 for neuron 0 and .129 for neuron 1), especially compared to the rest of the average cosines (respectively .397 and .431). These average percentages of low cosines appear significant (\(p(\chi^2) < .05; \chi^2_1 = 1064; \chi^2_2 = 1125\)) when evaluated inferentially using a \(\chi^2\) goodness-of-fit test taking as the theoretical distribution a 1\%/99\% distribution corresponding to a situation where these low cosines would be almost nonexistent (which should be the case if there were a relationship between categorical proximity and activation proximity). These elements are compatible with our hypothesis of categorical discontinuity of successive core-tokens postulating the existence of particularly low cosine similarities; a hypothesis, again, aimed at highlighting in a relatively extremized manner the fact that activation proximity (i.e., the level of categorical belongingness between two tokens) and cosine proximity (i.e., categorical proximity between these two tokens) do not go hand in hand (at the level of the overall distribution).

\begin{table}[H]  
\renewcommand{\arraystretch}{1.3}
\begin{adjustbox}{max width=\textwidth}
\begin{tabular}{p{0.48\textwidth}p{0.48\textwidth}} 
\includegraphics[height = 3.4cm]{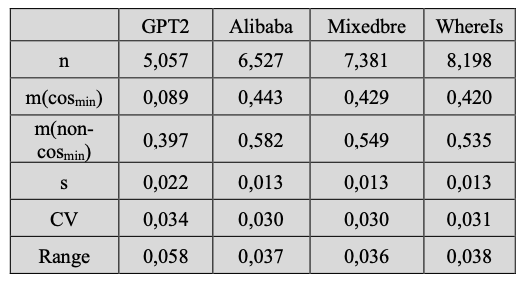}
\newline
\raggedright
\parbox{\linewidth}{\centering
\footnotesize
\textit{Table 8: Average Statistics of Low Cosine Similarity (cosine < min+0.1*Range) for Pairs of Successive Core-Tokens (Layer 0, n=6400).}} &
\includegraphics[height = 3.4cm]{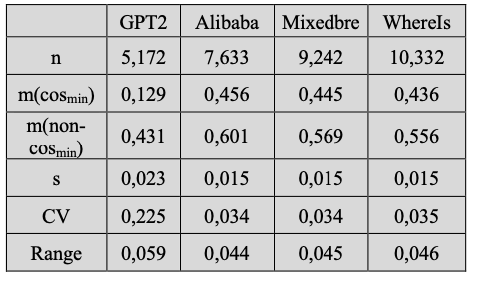}
\newline
\raggedright
\parbox{\linewidth}{\centering
\footnotesize
\textit{Table 9: Average Statistics of Low Cosine Similarity (cosine < min+0.1*Range) for Pairs of Successive Core-Tokens (Layer 1, n=6400).}} \\
\end{tabular}
\end{adjustbox}
\end{table}

\subsection{Categorical Inhomogeneity of Successive Core-Tokens at the Same Activation Level}

Continuing to explore the trend initially mentioned at the level of the overall distribution of cosines—that activation proximity is not equivalent to cosine proximity—we now test the following second hypothesis, which represents another perspective on this trend, again extremized to demonstrate the potential intensity it might have: there exists a categorical inhomogeneity of successive core-tokens at the same activation level. In other words, core-tokens having the same levels of activation are not categorically closest. This new extremized viewpoint this time focuses not immediately on the lowest cosine similarities, as was done before, but on cases where activations are close to being identical and should then be strongly associated with high cosine similarities if activation proximity and cosine proximity were phenomena that went hand in hand.

To operationalize the testing of this hypothesis, we define successive core-tokens with (almost) identical activations as those whose activation levels are equal to two decimal places. We define, for each neuron, a "d" distance indicator that equals the gap between the maximum cosine similarity of this neuron and the cosine similarity of successive core-tokens at the same activation; a distance whose superiority to the first quartile \(Q_1\) of the cosine distribution for this neuron is then verified, to show that statistically core-tokens at the same activation are not categorically closest. Tables No. 10 (layer 0) and No. 11 (layer 1) indicate first that tokens with (almost) identical activations are very numerous (respectively 47.25 and 35.31 tokens for 100 tokens per neuron), allowing us to consistently study the phenomenon of interest here. We see that the average distances are very large (.44 and .46, when measured with GPT-2XL embeddings, but also quite significant with other embeddings more inclined to over-represent strong cosine similarities). The percentages of neurons showing a distance d greater than \(Q_1\) are extremely high (respectively 80.72\% and 78.94\%); and significant (\(p(\chi^2) < .05\)) when evaluated inferentially with a univariate \(\chi^2\) goodness-of-fit test based on a theoretical 25\%/75\% distribution consistent with our use of \(Q_1\). These elements seem compatible with our hypothesis of mono-activational categorical inhomogeneity of successive core-tokens, at the level of the overall distribution of cosines.

\begin{table}[H]  
\renewcommand{\arraystretch}{1.3}
\begin{adjustbox}{max width=\textwidth}
\begin{tabular}{p{0.48\textwidth}p{0.48\textwidth}} 
\includegraphics[height = 2.7cm]{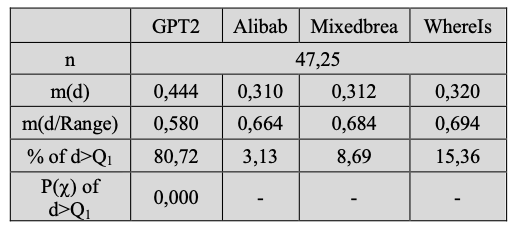}
\newline
\raggedright
\parbox{\linewidth}{\centering
\footnotesize
\textit{Table 10 : Average Statistics of Distance d (d = Max(COS(neuron)) - (COS(core-token(n), core-token(n')))) \& Comparison of d-Q1(cosine) for Pairs of Successive Core-Tokens with the Same Activation (Layer 0, n=6400).}} &
\includegraphics[height = 2.7cm]{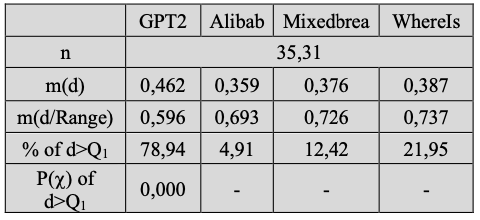}
\newline
\raggedright
\parbox{\linewidth}{\centering
\footnotesize
\textit{Table 11 : Average Statistics of Distance d (d = Max(COS(neuron)) - (COS(core-token(n), core-token(n')))) \& Comparison of d-Q1(cosine) for Pairs of Successive Core-Tokens with the Same Activation (Layer 1, n=6400).}} \\
\end{tabular}
\end{adjustbox}
\end{table}

To provide another angle of study on our hypothesis of categorical inhomogeneity, we implement the following complementary operationalization, consisting of comparing, for each neuron, its average cosine similarity relative to successive core-tokens at the same activation to the threshold of its third quartile (\(Q_3\)) of the cosine distribution, to show, in line with our hypothesis, the inferiority of the former to the latter. Tables No. 12 and No. 13 show that this is very much the case for all embedding systems (100\% with GPT-2XL embeddings). This is confirmed at the inferential level with extremely high percentages of cases where \(p(t) < .05\) in the context of a univariate Student's t-test comparison of means (cosines) to a standard (\(Q_3\)); again for all available embedding models; with 99.67\% for layer 0 and 95.66\% for layer 1 with GPT-2XL embeddings. We also note quite low average cosine similarities with measurements based on GPT-2XL embeddings (respectively .375 and .406), as well as with other embeddings (keeping in mind their tendency to overestimate strong cosine values). This second angle of view is again compatible, at the level of the overall distribution of cosines, with our hypothesis postulating that core-tokens having the same levels of activation are not categorically closest, intended to show in an exacerbated way to what extent activation proximity and cosine proximity would not be isomorphic elements; that is, to what extent the proximity of categorical belonging level between two tokens is a phenomenon that would be dissociated from the categorical proximity between these two tokens. Table No. 14 exemplifies this hypothesis of mono-activational categorical inhomogeneity qualitatively by showing pairs of tokens with the same activations that are widely disparate from a semantic viewpoint.

\begin{table}[H]  
\renewcommand{\arraystretch}{1.3}
\begin{adjustbox}{max width=\textwidth}
\begin{tabular}{p{0.48\textwidth}p{0.48\textwidth}} 
\includegraphics[width=\linewidth]{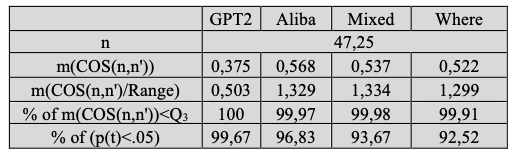}
\newline
\raggedright
\parbox{\linewidth}{\centering
\footnotesize
\textit{Table 12 : Average statistics of the cosines of successive core-token pairs with the same activation \& comparison to Q3(cosine) (Layer 0, n = 6400).}} &
\includegraphics[width=\linewidth]{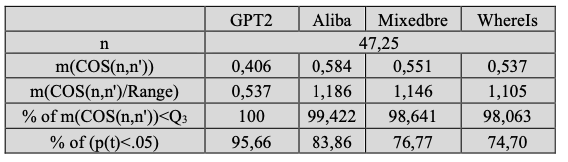}
\newline
\raggedright
\parbox{\linewidth}{\centering
\footnotesize
\textit{Table 13 : Average statistics of the cosines of successive core-token pairs with the same activation \& comparison to Q3(cosine) (Layer 1, n = 6400).}} \\
\end{tabular}
\end{adjustbox}
\end{table}

\begin{table}[H]  
\renewcommand{\arraystretch}{1.3}\centering
\includegraphics[width=\linewidth]{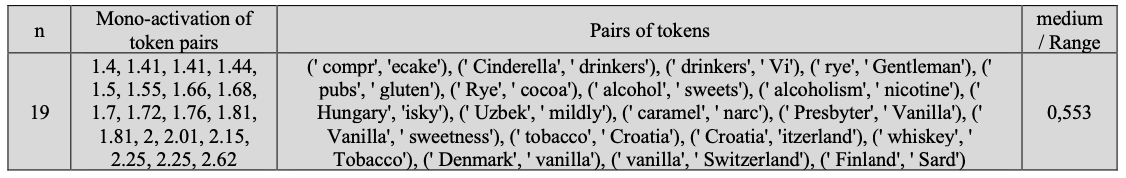}
\newline
\raggedright
\parbox{\linewidth}{\centering
\footnotesize
\textit{Table 14 : Examples of core-token pairs with the same activation level (Layer 1, neuron 113).}} 
\newline
\end{table}

\subsection{Synthesis}

  Let us now summarize our statistical treatments aimed at studying, at the level of successive core-tokens of each neuron, the characteristics of a possible existence of a relationship between activation proximity and cosine similarity (proximity). We obtained statistical results that are coherent with the two hypotheses formulated regarding phenomena of synthetic cognition:
\begin{itemize}
    \item A hypothesis of categorical discontinuity of successive core-tokens regarding their level of activation, positing that there are particularly low cosine similarities between successive core-tokens.
    \item A hypothesis of mono-activational categorical inhomogeneity of successive core-tokens, stating that core-tokens with the same levels of activation are not categorically closest.
\end{itemize}

\section{Discussion of Our Results}
Our series of studies questioned, concerning successive core-tokens, the possible existence of a relationship between proximity of categorical membership level and proximity of categorical similarity; proximities operationalized in terms of level of activation for the former and level of cosine similarity for the latter. Our results tend toward the idea of an independence, at the global level, between activation proximity and cosine proximity; i.e., towards a non-equivalence, for given tokens, between their proximity of belonging to a neuronal category and the intensity of their similarity. In other words, just because two tokens are close in terms of activation does not mean they are close at the categorical level. In the context of our current work, this phenomenology is manifested through two observables of synthetic cognition that we have attempted to elucidate: the categorical discontinuity of successive core-tokens and their mono-activational categorical inhomogeneity.

\subsection{Current Trends in Artificial Neuronal Explainability}

Several current interpretative trends in the field of synthetic neuronal explainability seem to us to be elementary keys, to be combined, for understanding the phenomenology of synthetic divergence between activation and similarity. We briefly present these trends below before attempting to elaborate on their interrelationship.

Synthetic neuronal polysemy is a primary concept. It refers to the idea that synthetic neurons are conceptually distributive (Fan et al., 2023), meaning they can correspond simultaneously to multiple semantic concepts (Bills et al., 2023). Thus, these authors suggest that neurons (i) may not have simple explanations but only long and disjunctive interpretations, (ii) should perhaps not be thought of as semantically homogeneous computational units. Bricken et al. (2023) further denote that formal neurons respond to unrelated traits.

Authors tend to link synthetic polysemy to the notion of superposition, which expresses the idea that cognitive properties and semantic traits can be ventilated within many polysemic neurons (Olah et al., 2020). And that a single concept is thus distributed across different neurons (Bills et al., 2023).

Another notion introduced by Bills et al. (2023), which seems relevant here and refers to a potential illusion of interpretability (Pichat, 2024b), is that of an alien concept. This means that formal neuronal concepts might be concepts for which humans have no word (no signifier in the sense of Saussure) or might even correspond to "natural abstractions" not yet discovered by humans (lacking human signified). This is insofar as, the authors indicate, language models deal with things different from us, for instance, statistical constructs useful for the task for which they have been trained, like predicting the next token.

Finally, Bricken et al. (2023) argue that the neuron does not constitute a good unit of semantic interpretation by introducing the idea of the existence of intermediate synthetic semantic vector spaces. A neural network would create a virtual intermediate vector space, each base vector of which would be an \textit{a priori} independent, fundamental, unique, and mono-semantic feature. Each of these features is obtained by linear combination of neurons, i.e., each feature is a vector on these neurons. Each feature thus constitutes an interpretable linear direction, an elementary semantic direction. Hence, the activation vector at the output of a neuronal layer could be decomposed in this intermediate space whose unit vectors are the elementary features. Each of these features would, by definition, be invisible at the level of a single neuron, which is why the neuron might not necessarily be the right unit of analysis according to the authors; they indicate, for example, in their study, that only 512 neurons can represent tens of thousands of features. From these fundamental semantic directions, more complex directions would be created, those constituted by the neurons, which then appear \textit{de facto} polysemic insofar as they are conceptually a compressed projection, i.e., low-dimensional, of these much vaster intermediate vector spaces.

\subsection{The Dissociation of Categorical Proximity and Categorical Similarity as a Sign of the Categorical Singularity of Synthetic Cognition}
Following our presentation of current trends in neuronal explainability, let's attempt to address our central empirical observation by reshaping and transposing them within a suitable explanatory framework. Why isn't activation proximity a corollary of similarity? Because a neuron codes for a synthetic category, which it creates in the context of its targeted activity, and this category is not unified; that is, it is polysemous. This polysemy makes this category appear as an alien concept (which it effectively is for our human cognition) insofar as it results from a superposition of sub-categorical dimensions generated by its intermediate categorical vector base (a base that we do not conceptualize in the same terms as Bricken et al. (2023) but rather, for a given neuron in layer n, in terms of categorical dimensions output from its precursor neurons in layer n-1). Two close activations (cf. the notion of categorical discontinuity) and even identical ones (cf. the notion of mono-activational categorical inhomogeneity) can thus correspond to crystallizations, materializations, local instantiations (by quantum analogy, we could speak of wave function collapses) of different sub-categorical dimensions. In other words, close activations can thus involve the actualizations of distinct sub-dimensions; which will then mechanically translate into cosine similarity measures demonstrating categorical discontinuity or inhomogeneity, concepts specific to synthetic cognition; at least concepts that appear to us when we study this synthetic cognition from our own human reference framework postulating an \textit{a priori} semantic logical coherence  between activation and similarity.

Neuronal polysemy does not seem to be categorically segmented into activation segments: categorical segments and activation segments appear to be two dissociated registers in synthetic neuronal cognition. Because this synthetic neuronal categorical cognition, unlike our human thought, is not categorically unified, at least not unified within concepts analogous to ours. Therefore, seeking a convergence between categorical proximity and cosine proximity is partly \textit{ipso facto} an anthropocentric approach that could only lead to the empirical correlational divergence that the synthetic notions of discontinuity and categorical inhomogeneity illustrate. This is especially true within the framework of our methodological approach of using cosine similarity as an instrument to measure categorical proximity: cosine similarity, being based on the vector categorical space of the initial embeddings of GPT-2XL (which is by construction more in tune with human semantics) is not the same as the recombinant vector spaces (by the values of their respective aggregation functions) of the neurons of the layers investigated.

Speaking of neuronal polysemy might epistemologically be a cognitive anthropocentrism (Pichat, 2024). Indeed, we use the term polysemy because synthetic categories appear semantically inhomogeneous given that we have no human thought categories to pair them with. But isn’t it the very nature of categorical abstraction to bring together initially separate categorical segments? The invoked polysemy is, in fact, the result of a categorical grouping to which we are not (at this stage of our conceptual evolution) accustomed, and we thus resemble in using this term the inhabitants of "flatland" from Watzlawick (1977) confined to viewing a multi-dimensional world through the only sub-dimensional prism proper to their world.

\section{Conclusion}

Our hypothesis of categorical discontinuity of successive core-tokens regarding their activation level (postulating that there are particularly low cosine similarities between successive core-tokens) and our hypothesis of mono-activational categorical inhomogeneity of successive core-tokens (stating that core-tokens with the same activation levels are not categorically closest) are complementary; this is because they pertain to the question of the global existence of a relationship between activation proximity and cosine proximity (similarity). They should be complemented by a third hypothesis, focusing on the possible evolution of the distributional dynamics of this relationship depending on the level of value of the activation segments; this hypothesis aims to investigate a possible categorical convergence of pairs of successive core-tokens based on activation, proposing that as the activation levels of successive core-tokens (i.e., close at the activation level) increase, the categorical variability of these core-tokens decreases (i.e., categorical proximity increases). We will soon publish our results in this area (Pichat et al., in press a), other results not mentioned here pushing us towards this new hypothesis formulated.

The key element posited, in our discussion of the obtained results in order to attempt to provide them with a coherent explanatory framework, is that the categorical segment created by a given neuron of a layer n (more precisely by its aggregation function among others) is \textit{de facto} decomposable into a vector space of sub-categorical dimensions; these being the result of a projection of the input vector space of this neuron, the input vector space being (by mathematical construction of its aggregation function) composed of the categorical output dimensions of each of the precursor neurons (on layer n-1) of this neuron. In other words, in terms of explainability, a neuron could immediately be thought of as multi-dimensional, that is, composed of sub-categorical dimensions which will tend to separately be involved depending on the tokens; at least the tokens with "low activation" (mono-categorical triggered) as opposed to the tokens with "high activations" which will tend to conjointly involve sub-dimensions (co-categorical triggered) as we are currently updating in some other research project. We will soon explore this postulate, in the context of a "genetic" study aimed at explaining the categorical abstraction carried out by artificial neurons on the basis of a reconstructive re-composition of the categorical segmentations of their most contributive precursor neurons (i.e., with the most significant neuronal connection weights).

\subsection*{Acknowledgements}

Michael Pichat thanks Sébastien Duizabo and David Abonneau (University Paris Dauphine PSL) for the stimulating symposiums and pedagogical experiments in applied AI they make possible with us, Emmanuel Brochier (IPC-Free Faculties of Philosophy and Psychology of Paris) for the exciting academic projects we have with him in AI, Igor Zatsman (Russian Academy of Sciences) and Nadia Buntman (Moscow State University Lomonosov) for the captivating AI and linguistic exchanges and Madeleine Pichat for her careful proofreading of this article.

\subsection*{Author Contributions}

Michael Pichat conceptualized and designed the study and served as the scientific lead. Enola Campoli contributed to various operational aspects of the study. William Pogrund handled data formatting and statistical processing. Michael Veillet-Guillem managed the SysAdmin part of the study. Jourdan Wilson participated in prompt engineering activities, translated the text into English, and formatted the published text. Judicael Poumay formated the initial data and advised about conceptual issues. Anton Melkoezrov was involved in prompt engineering activities and formatted the tables and diagrams. Samuel Demarchi provided advice on statistical studies. Armanouche Gasparian and Paloma Pichat made the realization of this study possible and scaffolded it at different levels.

\section*{Appendices}

\textbf{A.1 Descriptive Mean Satistics of Cosine Similarities of Pairs of Successive Core-Tokens (Control Neurons)}

\begin{table}[H]  
\renewcommand{\arraystretch}{1.3}\centering
\includegraphics[width=\linewidth]{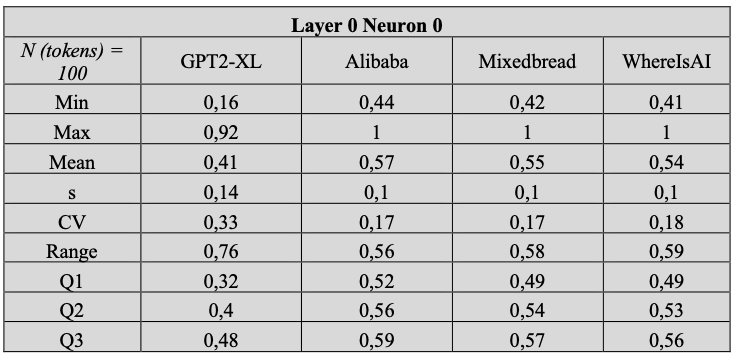}
\newline
\raggedright
\parbox{\linewidth}{\centering
\footnotesize
\textit{A.1 Descriptive Mean Statistics of Cosine Similarities of Pairs of Successive Core-Tokens (Layer 0, Neuron 0).}} 
\newline
\end{table}

\begin{table}[H]  
\renewcommand{\arraystretch}{1.3}\centering
\includegraphics[width=\linewidth]{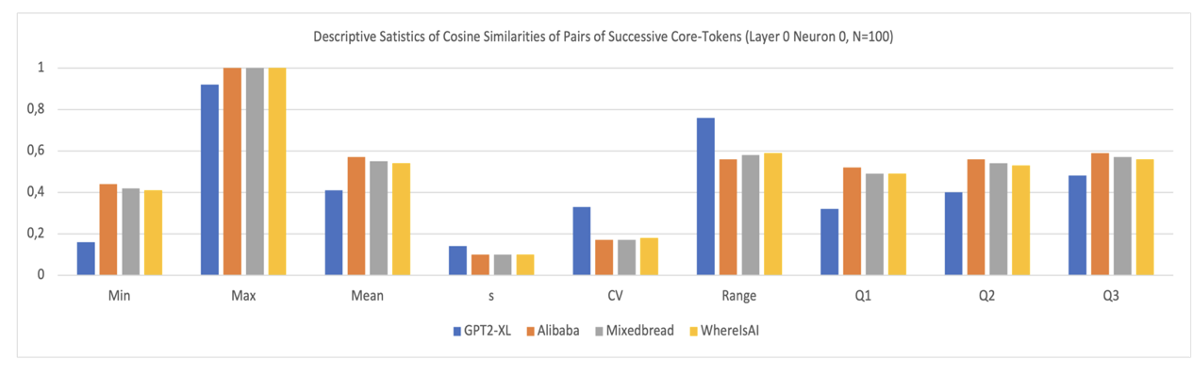}
\newline
\raggedright
\parbox{\linewidth}{\centering
\footnotesize
\textit{A.1 Descriptive Mean Statistics of Cosine Similarities of Pairs of Successive Core-Tokens (Descriptive Statistics of Cosine Similarities of Pairs of Successive Core-Tokens (Layer 0 Neuron 0, N=100).}} 
\newline
\end{table}

\begin{table}[H]  
\renewcommand{\arraystretch}{1.3}\centering
\includegraphics[width=\linewidth]{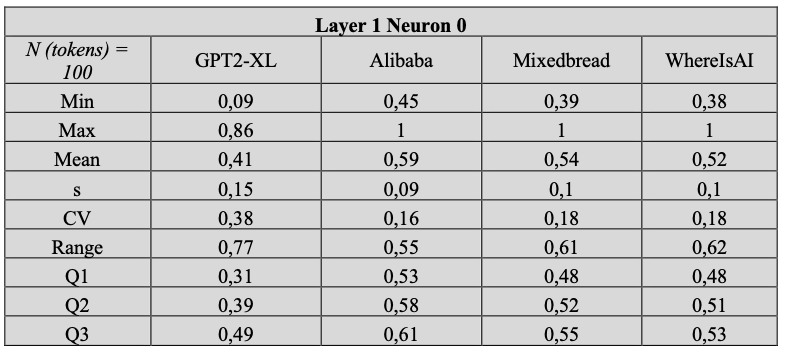}
\newline
\raggedright
\parbox{\linewidth}{\centering
\footnotesize
\textit{A.1 Descriptive Mean Statistics of Cosine Similarities of Pairs of Successive Core-Tokens (Layer 1, Neuron 0).}} 
\newline
\end{table}

\begin{table}[H]  
\renewcommand{\arraystretch}{1.3}\centering
\includegraphics[width=\linewidth]{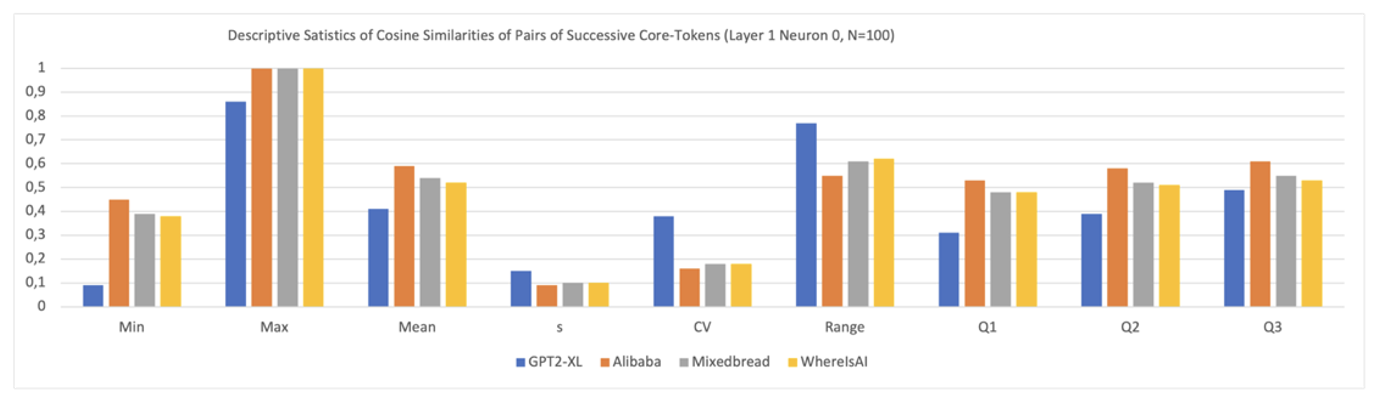}
\newline
\raggedright
\parbox{\linewidth}{\centering
\footnotesize
\textit{A.1 Descriptive Mean Statistics of Cosine Similarities of Pairs of Successive Core-Tokens (Descriptive Statistics of Cosine Similarities of Pairs of Successive Core-Tokens (Layer 1 Neuron 0, N=100)).}} 
\newline
\end{table}

\vspace{7cm} 

\vspace{2\baselineskip}
\textbf{A.2 Normality Ratio Statistics of Cosine Similarities of Pairs of Successive Core-Tokens (Control Neurons)}

\begin{table}[H]  
\renewcommand{\arraystretch}{1.3}\centering
\includegraphics[width=\linewidth]{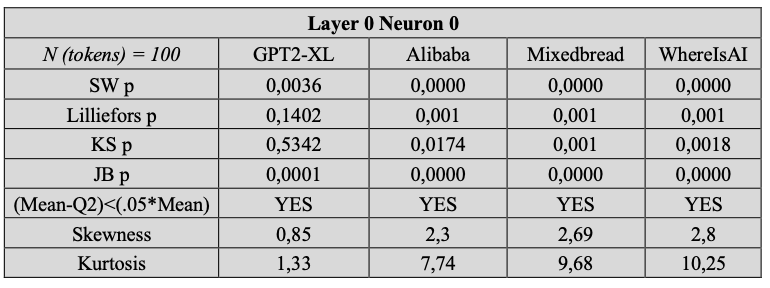}
\newline
\raggedright
\parbox{\linewidth}{\centering
\footnotesize
\textit{A.2 Normality Ratio Statistics of Cosine Similarities of Pairs of Successive Core-Tokens (Layer 0, Neuron 0).}} 
\newline
\end{table}

\begin{table}[H]  
\renewcommand{\arraystretch}{1.3}\centering
\includegraphics[width=\linewidth]{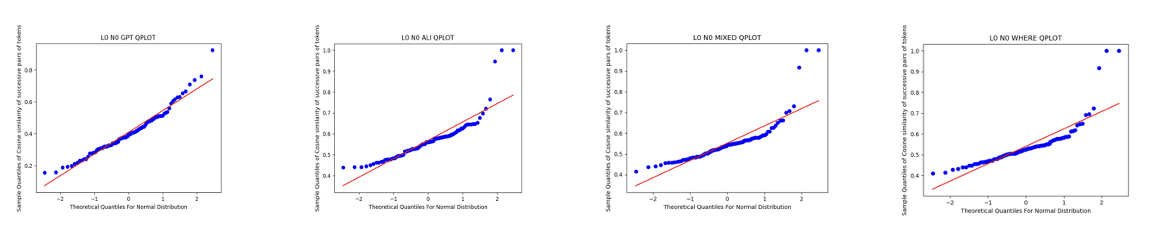}
\newline
\raggedright
\parbox{\linewidth}{\centering
\footnotesize
\textit{A.2 Normality Ratio Statistics of Cosine Similarities of Pairs of Successive Core-Tokens.}} 
\newline
\end{table}

\begin{table}[H]  
\renewcommand{\arraystretch}{1.3}\centering
\includegraphics[width=\linewidth]{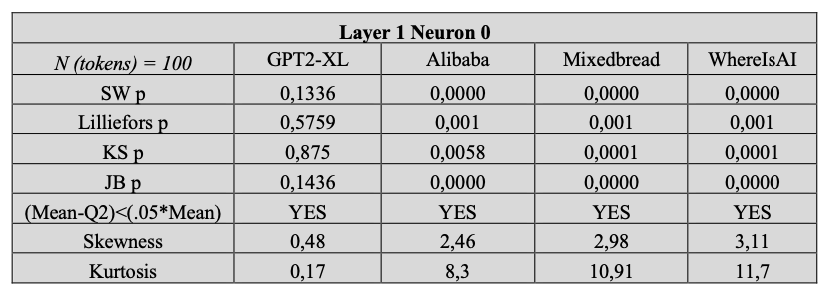}
\newline
\raggedright
\parbox{\linewidth}{\centering
\footnotesize
\textit{A.2 Normality Ratio Statistics of Cosine Similarities of Pairs of Successive Core-Tokens (Layer 1, Neuron 0).}} 
\newline
\end{table}

\begin{table}[H]  
\renewcommand{\arraystretch}{1.3}\centering
\includegraphics[width=\linewidth]{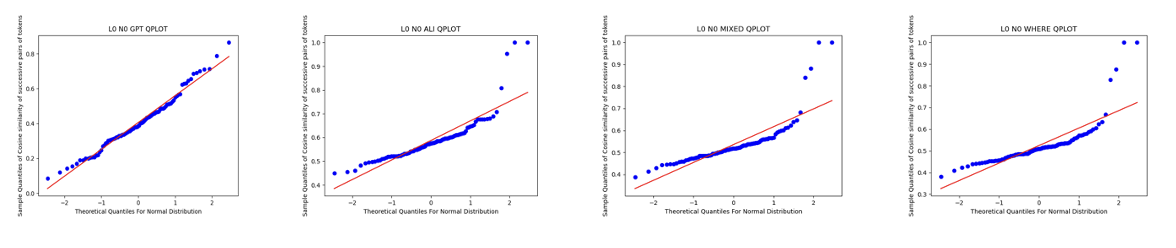}
\newline
\raggedright
\parbox{\linewidth}{\centering
\footnotesize
\textit{A.2 Normality Ratio Statistics of Cosine Similarities of Pairs of Successive Core-Tokens.}} 
\newline
\end{table}

\vspace{6\baselineskip}
\textbf{A.3 Distribution of Cosine Similarities of Pairs of Successive Core-Tokens as a Function of Activation Rank of the First Token of Pairs (Control Neurons)}

\begin{table}[H]  
\renewcommand{\arraystretch}{1.3}\centering
\includegraphics[width=\linewidth]{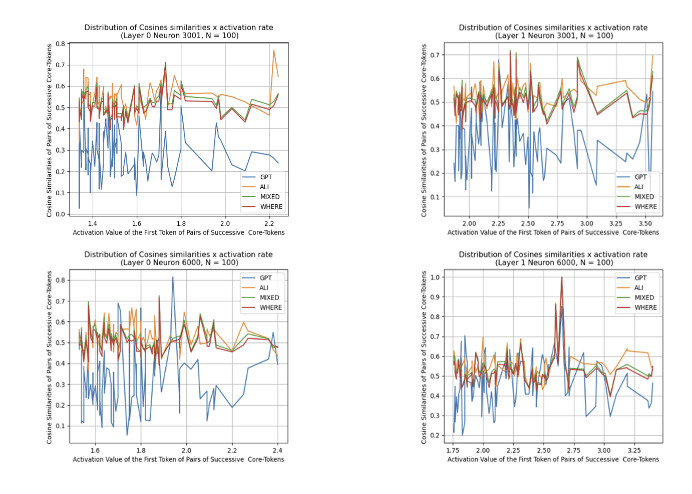}
\newline
\raggedright
\parbox{\linewidth}{\centering
\footnotesize
\textit{A.3 Distribution of Cosine Similarities of Pairs of Successive Core-Tokens as a Function of Activation Rank of the First Token of Pairs.}} 
\newline
\end{table}

\vspace{3\baselineskip}
\textbf{A.4} \textbf{Sample of Inferior Outliers of Cosine Similarities of Pairs of Successive Core-Tokens (Interquartile range, GPT2-XL)}

\begin{table}[H]  
\renewcommand{\arraystretch}{1.3}\centering
\includegraphics[width=\linewidth]{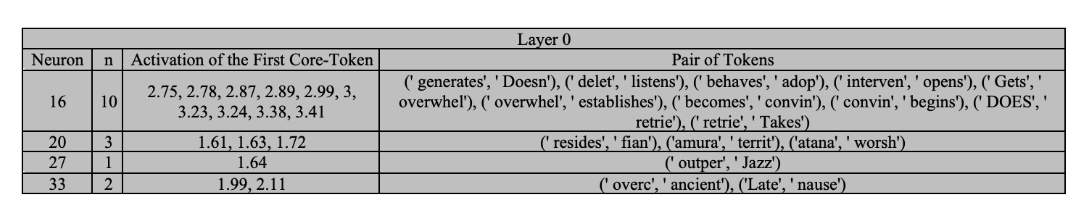}
\newline
\raggedright
\parbox{\linewidth}{\centering
\footnotesize
\textit{A.4 Sample of Inferior Outliers of Cosine Similarities of Pairs of Successive Core-Tokens (Interquartile range, GPT2-XL) (Layer 0).}} 
\newline
\end{table}

\begin{table}[H]  
\renewcommand{\arraystretch}{1.3}\centering
\includegraphics[width=\linewidth]{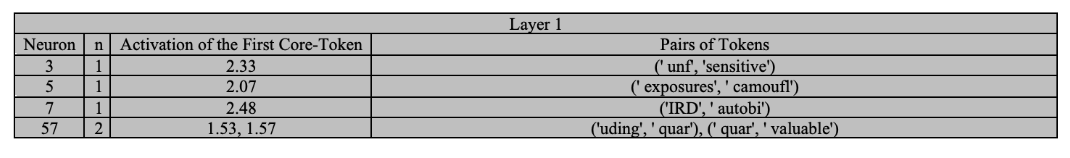}
\newline
\raggedright
\parbox{\linewidth}{\centering
\footnotesize
\textit{A.4 Sample of Inferior Outliers of Cosine Similarities of Pairs of Successive Core-Tokens (Interquartile range, GPT2-XL) (Layer 1).}} 
\newline
\end{table}

\vspace{4\baselineskip}
\textbf{A.5 Sample of Weak Cosine Similarities of Pairs of Successive Core-Tokens (GPT2-XL)}

\begin{table}[H]  
\renewcommand{\arraystretch}{1.3}\centering
\includegraphics[width=\linewidth]{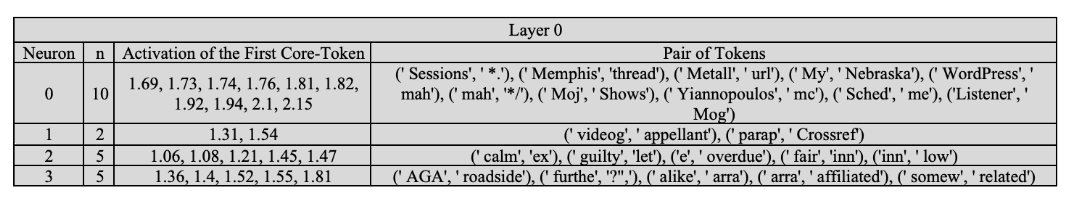}
\newline
\raggedright
\parbox{\linewidth}{\centering
\footnotesize
\textit{A.5 Sample of Weak Cosine Similarities of Pairs of Successive Core-Tokens (GPT2-XL) (Layer 0).}} 
\newline
\end{table}

\begin{table}[H]  
\renewcommand{\arraystretch}{1.3}\centering
\includegraphics[width=\linewidth]{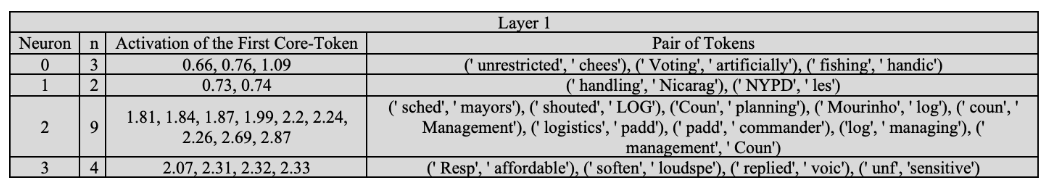}
\newline
\raggedright
\parbox{\linewidth}{\centering
\footnotesize
\textit{A.5 Sample of Weak Cosine Similarities of Pairs of Successive Core-Tokens (GPT2-XL) (Layer 1).}} 
\newline
\end{table}

\vspace{4\baselineskip}
\textbf{A.6 Sample of Successive Core-Tokens with Similar Activations}

\begin{table}[H]  
\renewcommand{\arraystretch}{1.3}\centering
\includegraphics[width=\linewidth]{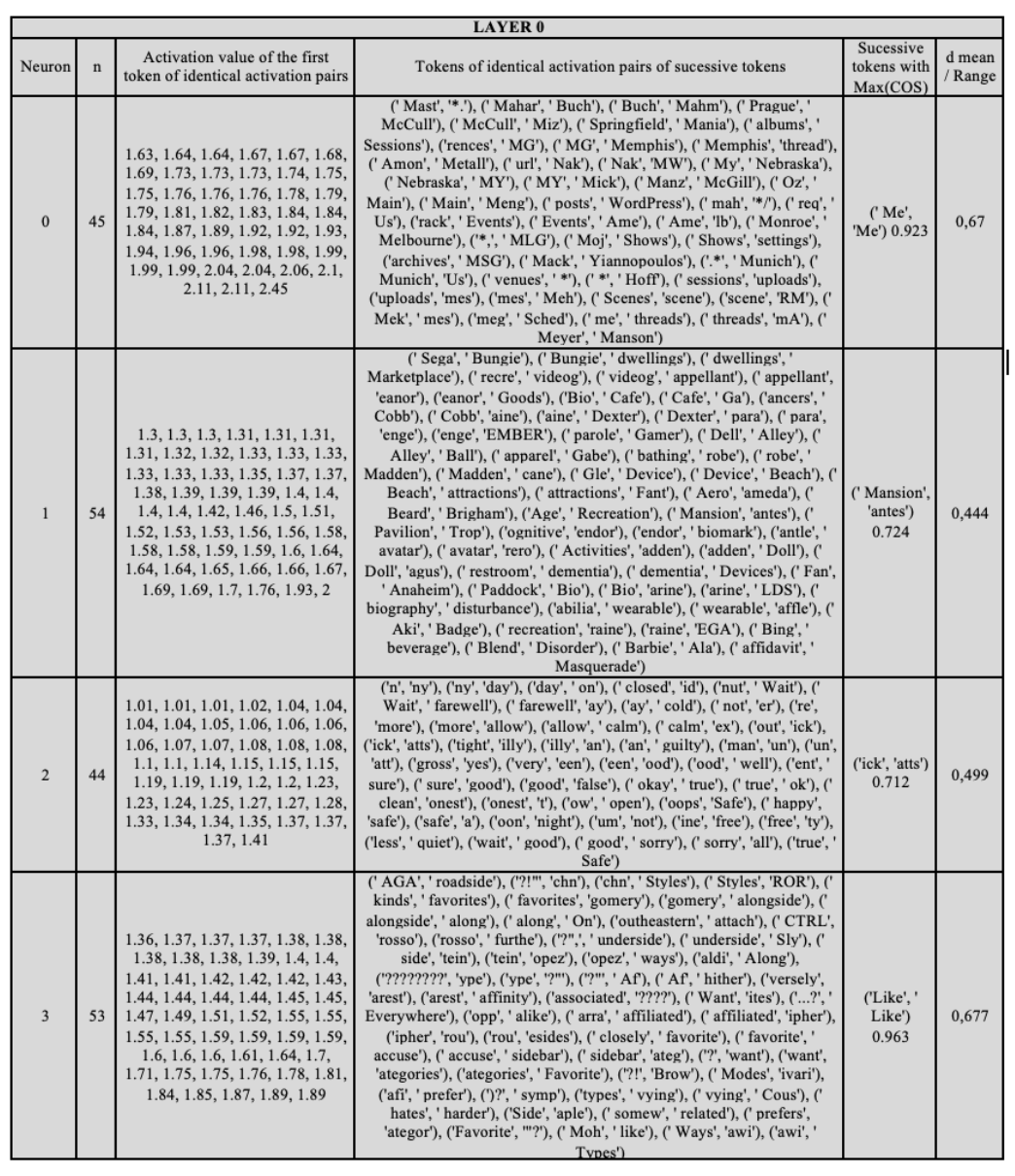}
\newline
\raggedright
\parbox{\linewidth}{\centering
\footnotesize
\textit{A.6 Sample of Successive Core-Tokens with Similar Activations (Layer 0).}} 
\newline
\end{table}

\begin{table}[H]  
\renewcommand{\arraystretch}{1.3}\centering
\includegraphics[width=\linewidth]{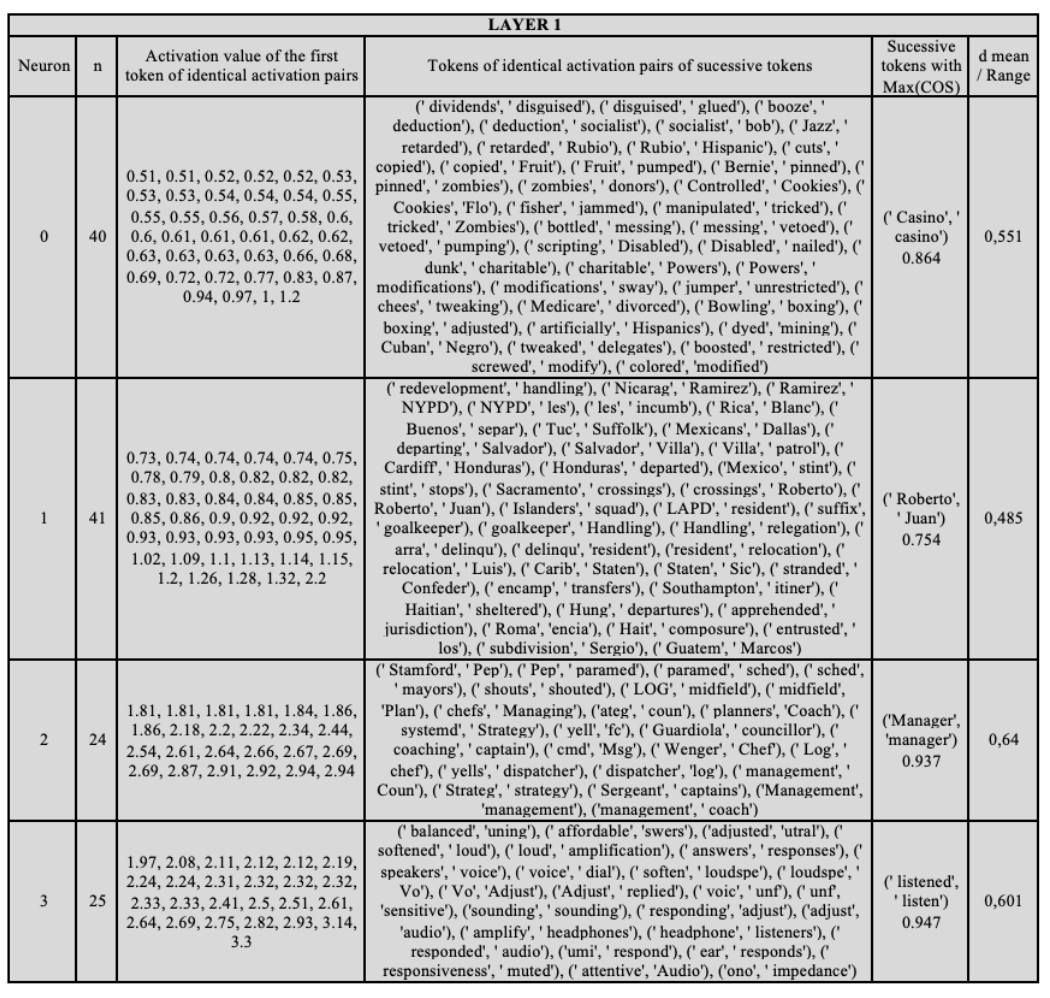}
\newline
\raggedright
\parbox{\linewidth}{\centering
\footnotesize
\textit{A.6 Sample of Successive Core-Tokens with Similar Activations (Layer 1).}} 
\newline
\end{table}

\vspace{4,3cm} 

\end{document}